\documentclass[preprint,nofootinbib,superscriptaddress]{revtex4-1}
\usepackage{amsmath,graphicx,hyperref}
\usepackage{color}
\usepackage{ctable}
\usepackage{dcolumn}   
\usepackage{bm}        
\usepackage{amssymb}   

\begin{document}

\title{Quarkonium polarization and the long distance matrix elements hierarchies using jet substructure}

\def\CMU{Department of Physics, Carnegie Mellon University, Pittsburgh, Pennsylvania, 15213, USA}
\def\Pitt{Pittsburgh Particle Physics Astrophysics and Cosmology Center (PITT PACC) \\ Department of Physics and Astronomy, University of Pittsburgh, Pittsburgh, Pennsylvania 15260, USA}

\author{Lin Dai}
\email[E-mail:]{lid33@pitt.edu}
\affiliation{\Pitt}

\author{Prashant Shrivastava}
\email[E-mail:]{prashans@andrew.cmu.edu}
\affiliation{\CMU}

\begin{abstract} \vspace{0.1cm}\baselineskip 3.0ex  
We investigate the quarkonium production mechanisms in jets at the LHC, using the Fragmenting Jet Functions (FJF) approach. Specifically, we discuss the jet energy dependence of the $J/\psi$ production cross section at the LHC. By comparing the cross sections for the different NRQCD production channels ($^1S_0^{[8]}$,$^3S_1^{[8]}$,$^3P_J^{[8]}$, and $^3S_1^{[1]}$), we find that at fixed values of energy fraction $z$ carried by the $J/\psi$, if the normalized cross section is a decreasing function of the jet energy, in particular for $z > 0.5$, then the depolarizing $^1S_0^{[8]}$ must be the dominant channel. This makes the prediction made in [Baumgart et al., JHEP {\bf 1411}, 003 (2014)] for the FJF's also true for the cross section. 
We also make comparisons between the long distance matrix elements extracted by various groups. This analysis could potentially shed light on the polarization properties of the $J/\psi$ production in high $p_T$ region.

\end{abstract}

\maketitle 
\baselineskip 3.3ex

\section{Introduction} 
Analyzing quarkonium production in jets provides a new way of probing the physics involved in their production. Recent developments include the LHCb measurements of $J/\psi$ production in jets \cite{Aaij:2017fak} and the related analyses~\cite{Ringer:2017, Bain:2017wvk, Belyaev:2017lbo}. A factorization theorem based on Non-Relativistic QCD (NRQCD)\footnote{\baselineskip 3.0ex NRQCD is an effective theory with a double expansion in the relative velocity $v$ of the heavy quark and anti-quark bound state and the strong coupling constant $\alpha_{s}$~\cite{Bodwin:1995,Caswell:1986,Brambilla:2000,Luke:2000,Rothstein:2017}.} can be used to calculate the cross section for $J/\psi$ production~\cite{Caswell:1986, Bodwin:1995}. Due to the large mass of the charm quark ($m_c$), the short distance production of the $c\overline{c}$ pair can be calculated perturbatively while the non-perturbative physics of the hadronization into a $J/\psi$ is captured by the long distance matrix elements (LDMEs) of the relevant production channels ($^1S_0^{[8]}$,$^3S_1^{[8]}$,$^3P_J^{[8]}$, and $^3S_1^{[1]}$). The predictive power of the theory is then predicated on our knowledge of these LDMEs. Different groups have extracted these matrix elements by using various fits to the data~\cite{Bodwin:2014gia,Butenschoen:2011yh,Butenschoen:2012qr,Chao:2012iv} but have arrived at very different values. Currently the NRQCD factorization theorem can consistently fit the unpolarized $J/\psi$ production cross section.

The $c\overline{c}$ pair produced by the fragmentation of a nearly on-shell gluon\footnote{\baselineskip 3.0ex For $J/\psi$ production via gluon fragmentation in NRQCD, the $^3S_{1}^{[1]}$ contribution is leading order in the $v$ expansion since the color octet channels are suppressed by $v^4$. But the $^3S_{1}^{[1]}$ is suppressed relative to the $^3S_{1}^{[8]}$ channel by power of $\alpha_{s}^2$. The matching onto $^3P_{J}^{[8]}$ and $^1S_{0}^{[8]}$ is down by $\alpha_{s}$ compared to $^3S_{1}^{[8]}$ but their LDMEs are of the same order as $^3S_{1}^{[8]}$ in $v$. An alternate power counting for charmonium production is formulated in Ref.~\cite{Fleming:2000}.} should inherit the transverse polarization of the gluon. Due to the spin symmetry of the leading order NRQCD Lagrangian, this polarization remains intact during the non-perturbative hadronization process (up to power corrections)~\cite{Cho:1995,Beneke:1996}. At leading order in $\alpha_{s}$, only the $^3S_{1}^{[8]}$ channel for the gluon contributes among the octet channels and since the color octet contribution is expected to dominate at high $p_T$~\cite{Beneke2:1996}, the $J/\psi$ meson should be produced with significant polarization at high $p_{T}$. However this prediction of NRQCD is at odds with the measurements of the $J/\psi$ polarization~\cite{Abulencia:2007us,Chatrchyan:2013cla,Aaij:2013nlm}. Understanding this polarization puzzle is one of the most important challenges in quarkonium physics~\cite{Brambilla:2011}.

A method based on jet substructure techniques to study the different production mechanisms of the $J/\psi$ was proposed in Ref.~\cite{Baumgart:2014}. By using the properties of the Fragmenting Jet Functions (FJF)~\cite{Stewart:2010}, it is predicted in Ref.~\cite{Baumgart:2014} that for a jet of energy $E$ and cone size $R$, containing a $J/\psi$ with energy fraction $z$ ($z={E_{J/\psi}}/{E}$), if the FJF is a decreasing function of the jet energy, then the dominant contribution to the $J/\psi$ production at high $p_T$ should be the depolarizing $^1S_0^{[8]}$ channel and hence, if confirmed by the data, this would resolve the polarization puzzle.

In this work, we investigate how the predictions of the diagnostic tool introduced in Ref.~\cite{Baumgart:2014} are affected by inclusion of the hard scattering effects. To do this, we calculate the total production cross section for the $J/\psi$. This should make the comparison of theory with experiments much simpler since the cross section can be directly measured. In order to make the distinction between various production channels, we calculate the cross section normalized in two different ways. In one case we normalize by summing over the contribution of all the channels and integrating over $z$ while in the other case we normalize by using the $1$-jet inclusive cross section. Additionally we also make comparisons between the LDMEs extracted by various groups. 

The main result of our paper is that the prediction made in Ref.~\cite{Baumgart:2014}, regarding the shapes of the FJF's, is also true for the cross section. By using a combination of differently normalized cross sections, we can break the degeneracy of the production channels and isolate the dominant contribution to the $J/\psi$ production at high $p_T$. Our results show that if the normalized cross section is a decreasing function of the jet energy at large $z$, in particular for $z>0.5$, then the $^1S_0^{[8]}$ channel dominates at high $p_T$ and this prediction should be easily verifiable with the LHC data. A recent work~\cite{Ringer:2017} also proposed using observables similar to ours to probe the $J/\psi$ production mechanisms.\footnote{\baselineskip 3.0ex Ref.~\cite{Ringer:2017} differentiates between the NRQCD global fits based on inclusive $J/\psi$ cross section and suggests using the polarization measurements of $J/\psi$ meson produced in the jets as a way of constraining the heavy quarkonium production mechanisms.}

\section{The Fragmenting Jet Functions}
\label{sec:FJF}
We briefly review the factorization theorem for the production of $J/\psi$~\cite{Stewart:2010,Liu:2011,Jain:2011,Jain:2012,Procura:2012,Jain:2013,Bauer:2013} before moving onto our main results in the next section. We consider the process $pp$ $\rightarrow$ dijets at $\sqrt{s} = 13$ TeV and integrate over one of the jets, assuming that the other jet contains an identified $J/\psi$. The dijet cross section~\cite{Stewart:2010} with one jet of energy $E$, cone size $R$ and a $J/\psi$ in the jet carrying an energy fraction $z$, is schematically of the form
\begin{equation}
\frac{d\sigma}{dEdz}=\sum_{a,b,i,j} H_{ab\rightarrow ij}\otimes f_{a/p}\otimes f_{b/p} \otimes J_{j} \otimes S \otimes \mathcal{G}^{\psi}_{i}(E,R,z,\mu),
\label{eq:master}
\end{equation}  
where $H_{ab\rightarrow ij}$ is the hard process, $f_{a/p}$ and $f_{b/p}$ are the parton distribution functions (PDF), $J_{j}$ is the jet function for the jet not containing the $J/\psi$, $S$ is the soft function and $\mathcal{G}^{\psi}_{i}(E,R,z,\mu)$ is the FJF for the jet containing the $J/\psi$. The parton $i$ can be a gluon, charm or an anti-charm (contributions of the other partons are suppressed). We are interested in the $E$ and $z$ dependence of the cross section, which comes from the hard function (including PDFs) and the FJF. We integrate over the jet originating from the parton $j$ so the jet function $J_{j}$ enters the cross section multiplicatively. The soft function $S$ does not affect $\mathcal{G}^{\psi}_{i}(E,R,z,\mu)$, $R$, $E$ and $z$ (up to power corrections)~\cite{Baumgart:2014} and so it also enters the cross section multiplicatively. Hence both the jet function $J_{j}$ and the soft function $S$ give an overall normalization to the cross section and are ignored in the rest of our analysis. In Ref.~\cite{Baumgart:2014}, the hard function was not included but here we calculate the normalized cross section, including both the charm quark and gluon contributions, and account for its $E$ dependence.

The FJF can be further factorized~\cite{Stewart:2010} into perturbatively calculable coefficients $\mathcal{J}_{ij}(E,R,z,\mu)$ and the fragmentation function $D_{j\rightarrow\psi}$: 
\begin{eqnarray}
\label{eq:FJF}
\mathcal{G}_{i}^{\psi}(E,R,z,\mu)=\int_{z}^{1}\frac{dy}{y}\mathcal{J}_{ij}(E,R,y,\mu)D_{j\rightarrow\psi}\Big(\frac{z}{y},\mu\Big)\Big(1+\mathcal{O}\Big(\frac{m^2_{\psi}}{4E^2\tan^2(R/2)}\Big)\Big).
\end{eqnarray}
The collection of NRQCD based fragmentation functions $D_{j\to \psi}$ used in this paper can be found in Ref.~\cite{Baumgart:2014}.

Large logarithms in $\mathcal{J}_{ij}(E,R,z,\mu)$ are minimized at the scale $\mu=2E\tan(R/2)(1-z)$ and can be easily resummed using the jet anomalous dimension~\cite{Procura:2012}. But we do not consider this resummation in this work since for us, $1-z \sim \mathcal{O}(1)$~\cite{Baumgart:2014}. Instead we evaluate the PDFs and $\mathcal{J}_{ij}(E,R,z,\mu)$ at the jet scale $\mu_{J}=2E\tan(R/2)$ and evolve the fragmentation function from $2m_{c}$ to the scale $\mu_{J}$ using the Dokshitzer-Gribov-Lipatov-Altarelli-Parisi (DGLAP) equation, 
\begin{equation}
\label{DGLAP}
\mu\frac{\partial}{\partial\mu}D_{i}(z,\mu)=\frac{\alpha_{s}(\mu)}{\pi}\sum_{j}\int_z^1 \frac{dy}{y}P_{i\rightarrow j}(z/y,\mu)D_{j}(y,\mu),
\end{equation}
where $P_{i\rightarrow j}(z/y,\mu)$ are the QCD splitting functions. We consider mixing between the charm quark and gluon splitting functions only for the $^3S_{1}^{[1]}$ channel.\footnote{\baselineskip 3.0ex The charm quark fragmentation into a $J/\psi$ is dominated by the $^3S_{1}^{[1]}$ channel because the color singlet and octet contributions start at same order in $\alpha_{s}$ but the color octet channels are suppressed in the $v$ expansion.} To leading order in $\alpha_{s}$, it can be shown that~\cite{Baumgart:2014}
\begin{equation}
\frac{\mathcal{G}_{i}^{\psi}(E,R,z,\mu_{J})}{2(2\pi)^3}\rightarrow D_{i\rightarrow\psi}(z,\mu_{J})+\mathcal{O}(\alpha_{s}(\mu_J)).
\end{equation}

Later in \ref{Norm1jetinclusive}, we will also consider the 1-jet inclusive cross section. This is calculated by replacing the FJF in Eq.~(\ref{eq:master}) with the jet function for a cone-type algorithm~\cite{ChrisLee:2010}. The FJFs are defined in Ref.~\cite{Stewart:2010} so that the sum over all possible fragmentations of a parton into hadrons equals the inclusive jet function.
\begin{equation}
J_{i}(E,R,\mu)=\frac{1}{2}\sum_{h}\int\frac{dz}{(2\pi)^3}z\mathcal{G}^{h}_{i}(E,R,z,\mu).
\label{eq:sumrule}
\end{equation}  
For further details about these calculations we refer the reader to Ref.~\cite{Baumgart:2014}. Throughout this paper we choose $m_{c}=1.4$ GeV and $R=0.4$.

\section{Discussion of the $J/\psi$ production mechanisms}
\label{sec:discussion}

In this section, we discuss the predictions for $J/\psi$ production in jets using the LDMEs extracted by various groups and reveal some generic features that are independent of these extractions. The LDMEs we use in this paper are summarized in Table \ref{tb-ldme}. Refs.~\cite{Butenschoen:2011yh,Butenschoen:2012qr} use a global fit to 194 data points from 26 data sets and predict significant polarization of the $J/\psi$ in the high $p_T$ region, which contradicts the measurements at the Tevatron~\cite{Abulencia:2007us} and the LHC~\cite{Chatrchyan:2013cla, Aaij:2013nlm}. The extractions in Refs.~\cite{Bodwin:2014gia, Chao:2012iv} focus on the high $p_T$ region and attempt to solve the polarization puzzle.
\begin{table*}[t]
\begin{center}
\begin{tabular}{|r||r|r|r|r|}
\hline
 &  $\langle \mathcal{O}^{J/\psi}(^3S_1^{[8]}) \rangle $ & $\langle \mathcal{O}^{J/\psi}(^1S_0^{[8]}) \rangle$ & $\langle \mathcal{O}^{J/\psi}(^3P_0^{[8]}) \rangle/m_c^2 $ & $\langle\mathcal{O}^{J/\psi}(^3S_1^{[1]}) \rangle$ \\
 & $\times 10^{-2}$ GeV$^3$ &  $\times 10^{-2} $GeV$^3$ & $\times10^{-2} $GeV$^3$& $\times$ GeV$^3$ \\
\hline 
Bodwin et al. Ref.~\cite{Bodwin:2014gia} & $1.1\pm 1.0$ & $9.9 \pm 2.2$ & $0.56 \pm 0.51$ & $1.32$\\ 
\hline
Butenschoen et al. Ref.~\cite{Butenschoen:2011yh,Butenschoen:2012qr} &  $0.224 \pm 0.059$ & $4.97 \pm 0.44$ & $-0.82 \pm 0.10$ & $1.32$\\ 
\hline
Chao et al. Ref.~\cite{Chao:2012iv} &  $0.30 \pm 0.12$ & $8.9\pm 0.98$ & $0.56 \pm 0.21$ & $1.16$  \\
\hline
\end{tabular}
\end{center}
\caption{LDMEs extracted by various groups used in this paper.}
\label{tb-ldme}
\end{table*}

\subsection{Normalized $J/\psi$ production cross section}
\label{sec:ldmedependent}

\begin{figure*}[t]
	\begin{center} 
		\includegraphics[width=13cm]{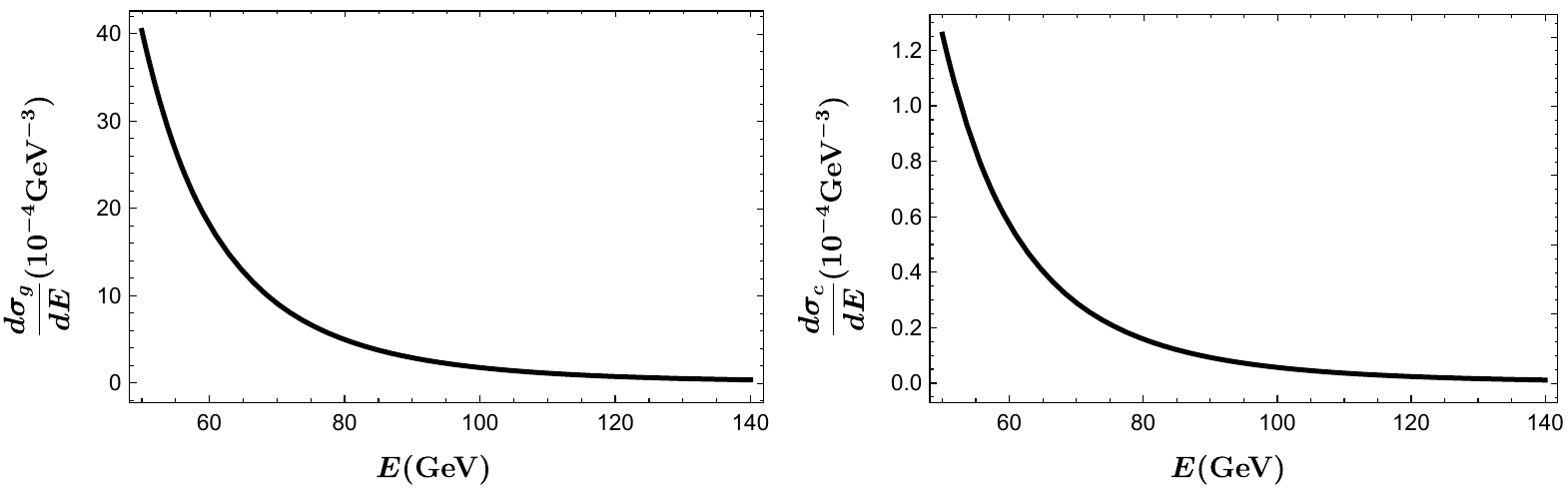}
	\end{center}
	\vspace{-0.3cm}
	\caption{\baselineskip 3.0ex Cross sections for inclusive gluon and charm jets at the LHC. The center of mass energy is $\sqrt{s}=13$ TeV.}
	\label{plotgc}
\end{figure*}

To discuss the dependence of $J/\psi$ production on the associated jet energy, we use a normalized differential cross section defined as
\begin{equation}
\label{eq:norm}
\frac{d\tilde{\sigma}_i}{dEdz}\equiv {\frac{d\sigma_i}{dEdz}}\Bigg/{\sum_i\int_{z_{min}}^{z_{max}}\mathrm{d}z\,\frac{d\sigma_i}{dEdz}},
\end{equation}
and
\begin{equation}
\label{eq:norm2}
\frac{d\tilde{\sigma}}{dEdz}\equiv \sum_i \frac{d\tilde{\sigma}_i}{dEdz},
\end{equation}
where $i$ denotes different $J/\psi$ production channels (i.e., for the gluon initiated jets $i \in$ \{$^1S_0^{[8]}$, $^3S_1^{[8]}$, $^3P_J^{[8]}$, $^3S_1^{[1]}$\} and for the charm initiated jets $i={^3}S_1^{[1]}$), and $d\sigma_i/dEdz$ is defined in Eq.~(\ref{eq:master}). 

In Eq.~(\ref{eq:norm}), $z_{min}~(z_{max}$) should not be too close to $0~(1)$ where the factorization breaks down. The motivation for studying this normalized cross section is that we want to isolate the properties of quarkonium fragmentation in jets from the hard process that generates the jet initiating parton's. Fig.~(\ref{plotgc}) shows the energy distributions of the hard process for gluon and charm jets at the LHC\footnote{\baselineskip 3.0ex We consider leading order partonic cross sections convoluted with PDF~\cite{Collider, MSTW:2008nlo}, which includes the following processes: $gg\rightarrow gg$, $gq(\overline{q}) \rightarrow gq(\overline{q}) $, $q\overline{q} \rightarrow gg$, $gg \rightarrow c\overline{c}$, $gc(\overline{c}) \rightarrow gc(\overline{c})$, $cc \rightarrow cc$, $\overline{c}~\overline{c} \rightarrow \overline{c}~\overline{c}$, $cq(\overline{q}) \rightarrow cq(\overline{q})$, $\overline{c}q(\overline{q}) \rightarrow \overline{c}q(\overline{q})$, $q\overline{q} \rightarrow c\overline{c}$, $c\overline{c} \rightarrow c\overline{c}$.}. For all the figures in this paper, we fix the center of mass energy to be $\sqrt{s}=13$ TeV. 

\begin{figure*}[t]
	\begin{center} 
		\includegraphics[width=17cm]{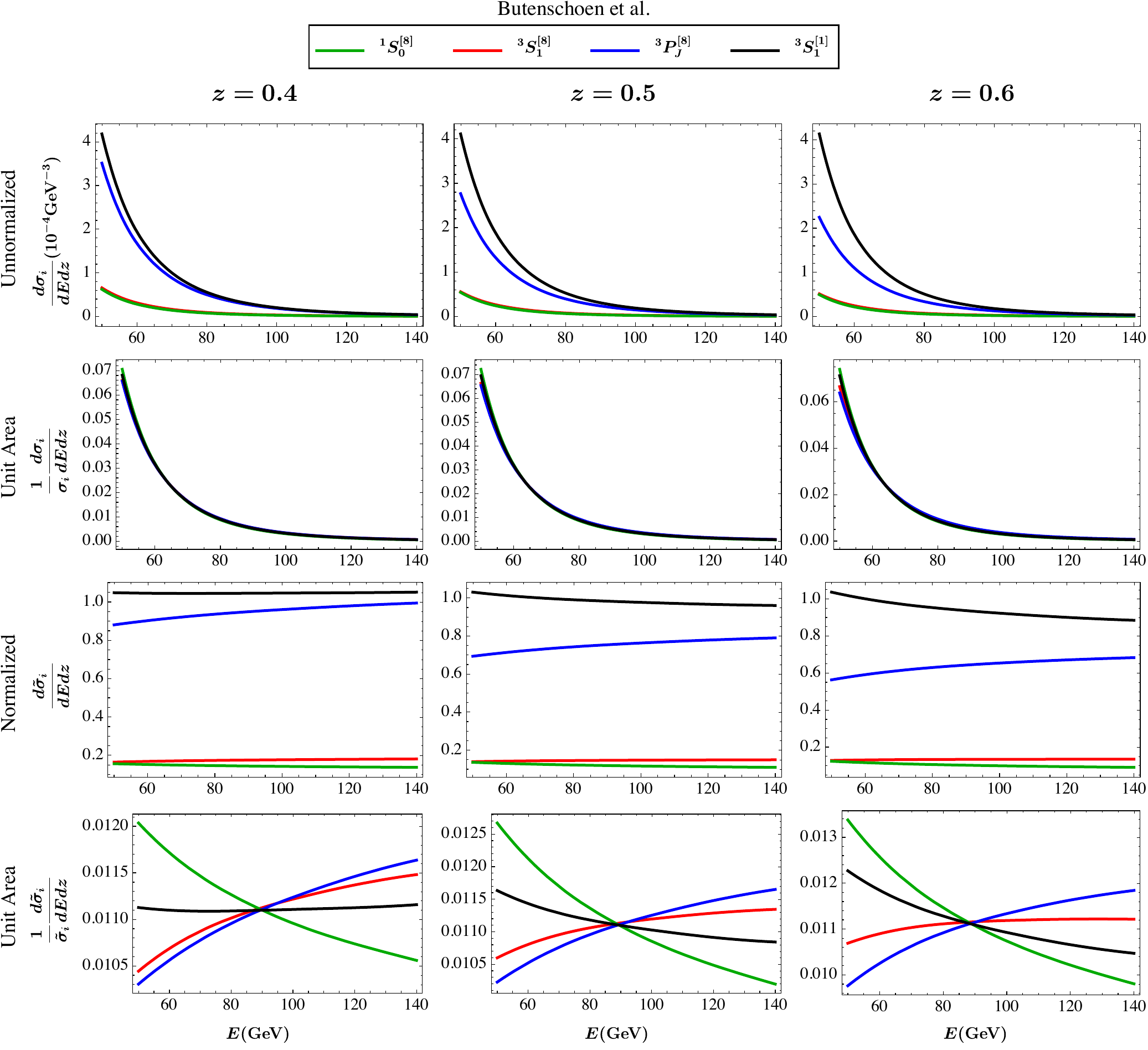}
	\end{center}
	\vspace{-0.3cm}
	\caption{\baselineskip 3.0ex Cross sections for the different production channels at $z=0.4$, $0.5$, and $0.6$ as a function of the jet energy. The first two rows show the unnormalized cross sections ($d\sigma_i/dEdz$), with the second row showing plots normalized to unit area for a better visualization of the shapes, i.e., we multiply each curve of the first row by an appropriate constant to get the corresponding curve in the second row.  Similar plots for the normalized cross section ($d\tilde{\sigma}_i/dEdz$) are shown in the third and fourth row. The LDMEs are from Butenschoen et al.'s extractions~\cite{Butenschoen:2011yh}.}
	\label{plotCompNorm}
\end{figure*}

Fig.~(\ref{plotCompNorm}) shows the comparison of the normalized (Eq.~(\ref{eq:norm})) and unnormalized cross sections (Eq.~(\ref{eq:master})), where the LDMEs from Ref.~\cite{Butenschoen:2011yh, Butenschoen:2012qr} are used with $z_{min}=0.3$ and $z_{max}=0.8$. Corresponding plots for the LDMEs of Ref.~\cite{Bodwin:2014gia} and Ref.~\cite{Chao:2012iv} are shown in appendix \ref{Bodwin} and \ref{Chao} respectively. We would like to emphasize the fact that both the unnormalized and normalized cross sections are directly measurable in experiments, although the normalized cross section has a better resolving power than the unnormalized cross section. In particular, the unnormalized cross section is a decreasing function of $E$ for all the production channels due to the decreasing nature of the hard process, while the normalized cross section can be an increasing function for certain production channels due to the properties of their FJF's. 

A measurement of the normalized cross section (Eq.~(\ref{eq:norm})) for $z>0.5$, can help identify both the dominant channel and the favored set of LDMEs. From Fig.~(\ref{plotCompNorm}), we can see that if $d\tilde{\sigma}_i/dEdz$ turns out be a decreasing function of the jet energy for $z>0.5$, then the depolarizing $^1S_{0}^{[8]}$ should be the dominant channel. We find this result to be true for LDME extractions of Ref.~\cite{Bodwin:2014gia} as well (see appendix \ref{Bodwin}).      
  
\begin{figure*}[t]
	\begin{center} 
		\includegraphics[width=17cm]{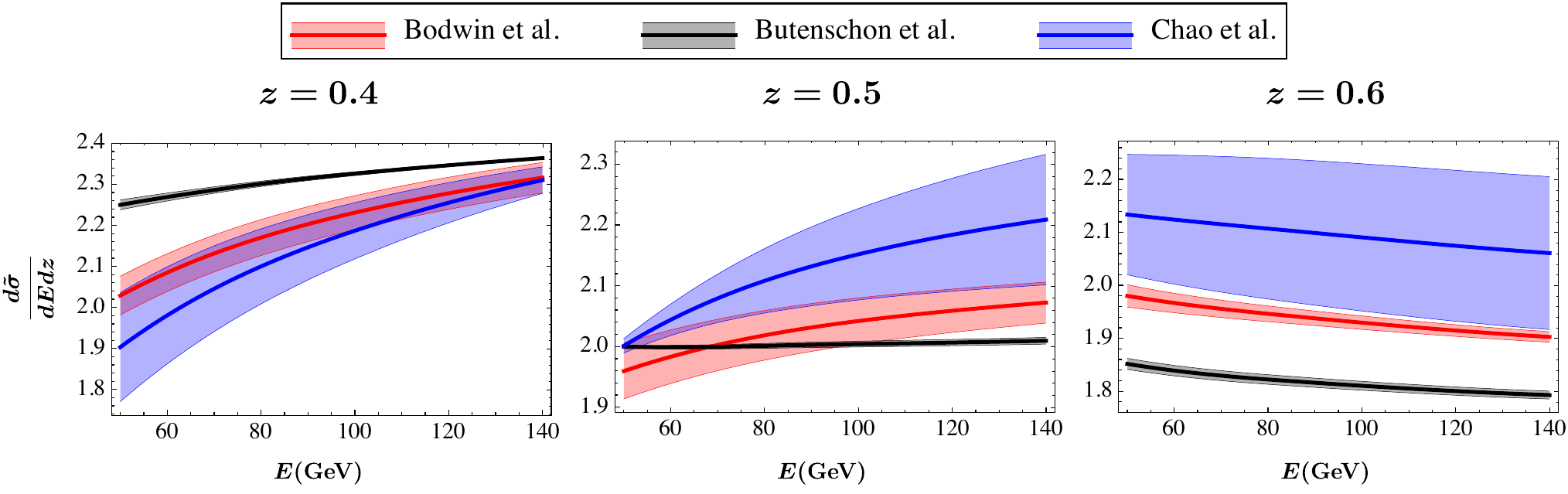}
		\end{center}
		\vspace{-0.3cm} 
		\caption{\baselineskip 3.0ex Total normalized cross section (i.e. $d\tilde{\sigma}/dEdz$ defined in Eq.~(\ref{eq:norm2})) with error bands. Red, black, and blue curves correspond to Bodwin et al.~\cite{Bodwin:2014gia}, Butenschoen et al.~\cite{Butenschoen:2011yh,Butenschoen:2012qr}, and Chao et al.'s~\cite{Chao:2012iv} extractions, respectively.}
		\label{plotCombErr}
\end{figure*}

In Fig.~(\ref{plotCombErr}), we show the jet energy dependence of the total normalized cross sections (Eq.~(\ref{eq:norm2})) based on different LDME extractions. The error bands are purely due to the LDME uncertainties, that is, we consider the uncertainty due to each LDME and sum by quadrature to obtain the total uncertainties\footnote{\baselineskip 3.0ex To obtain the error bands corresponding to the extraction from Bodwin et al., we have used the error correlation matrix not shown in the original paper~\cite{Bodwin:private}.}. It can be seen in Fig.~(\ref{plotCombErr}) that as $z$ goes from $0.4$ to $0.6$, the shapes change from an increasing function to a decreasing function. However since different extractions have distinct slopes, this observable has the potential power to test these extractions at the LHC. A different choice of $(z_{min},z_{max})$ does not change our arguments as we demonstrate in appendix \ref{NormalizationInsensitivity}.

We also consider the possibility that for $z>0.3$, the contribution of the $^3S_{1}^{[1]}$ channel to the $J/\psi$ production is negligible for the $p_T$ range considered here~\cite{Brambilla:2011,Beneke2:1996,Bodwin:2014gia}. We test this by ignoring the $^3S_1^{[1]}$ channel contribution to the  normalization and arrive at the same conclusion of $^1S_{0}^{[8]}$ being the dominant contribution if the normalized cross section decreases with jet energy for $z>0.5$ (see appendix \ref{ColorOctet}).
\subsection{Normalization using 1-jet inclusive cross section}
\label{Norm1jetinclusive} 
We now normalize the cross section in such a way that the denominator is independent of the LDMEs. This allows us to make a direct comparison of our results to those of Ref.~\cite{Baumgart:2014}. The normalization is defined as 
\begin{equation}
\label{eq:normjc}
\frac{d\hat{\sigma}_i}{dEdz}\equiv {\frac{d\sigma_i}{dEdz}}\Bigg/{\frac{d\sigma_J}{dE}},
\end{equation}
and
\begin{equation}
\label{eq:normjc2}
\frac{d\hat{\sigma}}{dEdz}\equiv\sum_i \frac{d\hat{\sigma}_i}{dEdz},
\end{equation}
where $d\sigma_i/dEdz$ is the same as that in Eq.~(\ref{eq:master}) and $d\sigma_J/dE$\footnote{\baselineskip 3.0ex This includes the contributions of gluon, light quarks, charm and bottom jets.} is the $1$-jet inclusive cross section\footnote{\baselineskip 3.0ex The definition of Eq.~(\ref{eq:normjc}) is essentially the same as the jet fragmentation function introduced in Ref.~\cite{Ringer:2017}, except that we have integrated the jet pseudorapidity over the region $|\eta_J|<1.2$ for the denominator and numerator.}. Note that the $z$-dependence of Eq.~(\ref{eq:normjc}) comes only from the $\mathcal{G}_i^{J/\psi}(E,R,z,\mu)$ in Eq.~(\ref{eq:master}). 
\begin{figure*}[t]
	\begin{center} 
		\includegraphics[width=17cm]{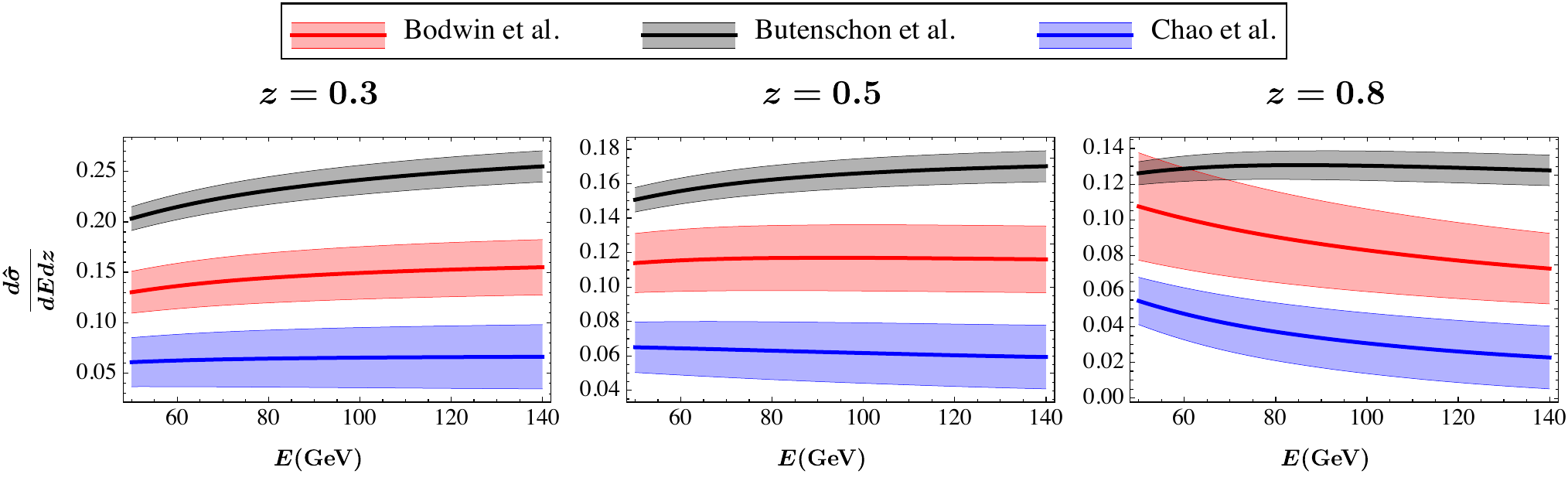}
	\end{center}
	\vspace{-0.3cm}
		\caption{\baselineskip 3.0 ex Total normalized cross section (i.e. $d\hat{\sigma}/dEdz$ defined in Eq.~(\ref{eq:normjc2})) with error bands. Red, black, and blue curves correspond to Bodwin et al.~\cite{Bodwin:2014gia}, Butenschoen et al.~\cite{Butenschoen:2011yh,Butenschoen:2012qr}, and Chao et al.'s~\cite{Chao:2012iv} extractions, respectively.}
		\label{plotCombErrjcn}
\end{figure*}

\begin{figure*}[t]
	\begin{center} 
		\includegraphics[width=17cm]{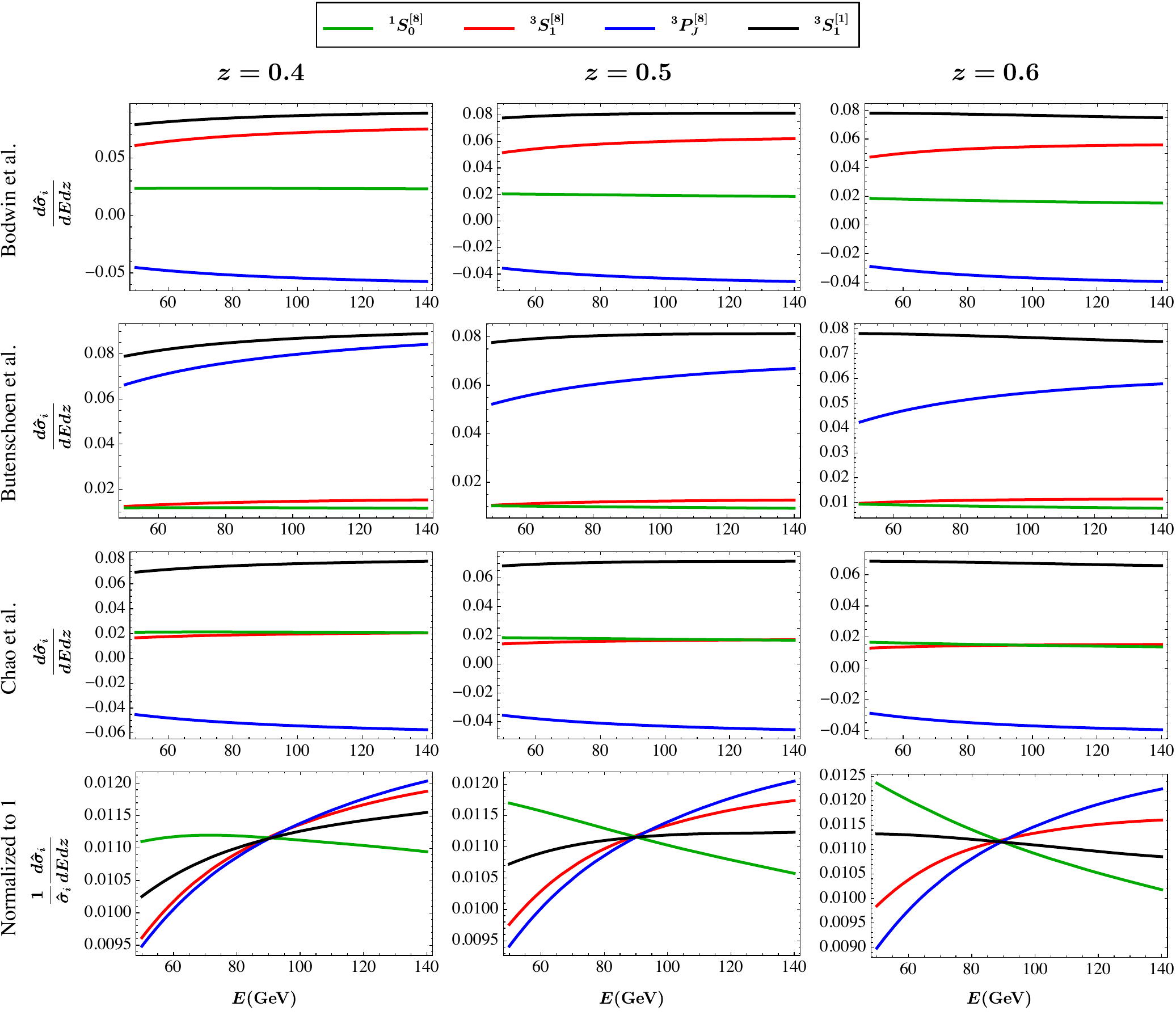}
	\end{center}
	\vspace{-0.3cm}
		\caption{\baselineskip 3.0ex Comparisons of the production channels for various LDMEs using Eq.~(\ref{eq:normjc}). Last row shows the plots normalized to unit area. This is indicated by $1/\hat{\sigma}_i$ for the cross section label in the fourth row, which also cancels the LDME dependence of the numerator.}
		\label{plotJetNorm}
\end{figure*}

Fig.~(\ref{plotCombErrjcn}) shows the total $J/\psi$ production cross section based on Eq.~(\ref{eq:normjc2}). The key feature of this plot is that the arguments given Ref.~\cite{Baumgart:2014} based on the FJFs are also true for the cross section (see Fig. 6 in Ref.~\cite{Baumgart:2014})\footnote{\baselineskip 3.0ex To facilitate direct comparison of our Fig.~(\ref{plotCombErrjcn}) to Fig.~(6) in Ref.~\cite{Baumgart:2014}, we make plots for $z=0.3, 0.5$ and $0.8$.}.  Specifically, when $z>0.5$, the shapes of the curves are very different for the extraction based on a global fit (black curves) and the other two based on fit to high $p_T$ region (red and blue curves). Since the extractions from the global fit and high $p_T$ fit give rise to different slopes for the $J/\psi$ production cross section, one can test which set of the LDME extractions are preferred by measuring these slopes. Note that because our results are for the cross section, all the curves have positive values, in contrast to the gluon FJF for the LDMEs of Ref.~\cite{Chao:2012iv} (shown in Fig.~(6) of Ref.~\cite{Baumgart:2014}) which became negative at large energies.

 
 In Fig.~(\ref{plotJetNorm}), we plot the $E$ dependence of the individual $J/\psi$ production channels for the different LDMEs using Eq.~(\ref{eq:normjc}). We find that if the measurements of the observable defined in Eq.~(\ref{eq:normjc}) results in a cross section which is a decreasing function of the jet energy for $z>0.5$, then the $^1S_0^{[8]}$ channel should have an anomalously large contribution to the $J/\psi$ production. The fourth row in Fig.~(\ref{plotJetNorm}), with the curves normalized to unit area, clearly shows that only $^1S_0^{[8]}$ channel is a decreasing function of jet energy for $z>0.5$.  
 Hence a verification of our results in this and the previous section will give strong evidence in favor of the depolarizing $^1S_0^{[8]}$ channel being dominant at high $p_T$ and provide a clear explanation for the lack of polarization in the $J/\psi$ production at high $p_T$. Note that in the fourth row of Fig.~(\ref{plotJetNorm}), the LDME dependence gets canceled due to normalization to unit area and so the prediction for $^1S_0^{[8]}$ channel being dominant at high $p_T$ is independent of any specific LDME extractions.

To conclude this section, we mention a few things about the normalization conventions in Eq.~(\ref{eq:norm}) and Eq.~(\ref{eq:normjc}). First of all, both the normalizations can be directly tested in experiments. Also since both the numerator and denominator of Eq.~(\ref{eq:norm}) depend on the LDMEs, they are statistically correlated and hence the width of error bands in Fig.~(\ref{plotCombErr}) is reduced. However, Eq.~(\ref{eq:normjc}) does not have such a correlation since the jet cross section used for the normalization is independent of the LDMEs. Indeed, if we look at Bodwin et al.'s extraction near $z=0.5$ and $E=100$ GeV, the ratio of the width of error band to the center value is $\sim 4\%$ in Fig.~(\ref{plotCombErr}) and $\sim 30\%$ in Fig.~(\ref{plotCombErrjcn}). On the other hand, in both Fig.~(\ref{plotCombErr}) and Fig.~(\ref{plotCombErrjcn}), the shapes of blue and red curves (high $p_T$ fit) are in contrast to the black curve (global fit).  

\section{Conclusion}
\label{conc}

In this paper, we have looked at the total cross section for $J/\psi$ production at the LHC by using the FJF approach. We make comparisons between the different NRQCD production channels for the $J/\psi$. We show that if for $z>0.5$ the normalized cross section is a decreasing function of jet energy, then the depolarizing $^1S_0^{[8]}$ should be the dominant production channel at high $p_T$. We find this to be true for two sets of normalized cross sections. Our results confirm that the prediction made in Ref.~\cite{Baumgart:2014} regarding the decreasing nature (with $E$) of the FJF for $^1S_0^{[8]}$ channel, does not change by inclusion of the hard scattering effects. Using our normalized cross sections, one can also test which set of the LDME extractions are favored.

\acknowledgments
The authors would like to thank Adam Leibovich and Ira Rothstein for their guidance and comments on the manuscript. We would also like to thank James Russ for useful discussions. LD was supported in part by NSF grant PHY-1519175.
\clearpage
\appendix
{
\section{Unnormalized and Normalized cross sections for Bodwin et.al}
\label{Bodwin} 
Fig.~(\ref{plotCompNormBO}) shows the unnormalized (Eq.~(\ref{eq:master})) and normalized cross section (Eq.~(\ref{eq:norm})) for Bodwin et al.'s LDME extractions~\cite{Bodwin:2014gia}. The $^3P_{J}^{[8]}$ channel contribution is negative, which is a feature of these LDMEs as it leads to a cancellation between the $^3S_{1}^{[8]}$ and $^3P_{J}^{[8]}$ channels, making the depolarizing $^1S_{0}^{[8]}$ the dominant production channel of $J/\psi$ for $z>0.5$.
\begin{figure*}[htbp!]
	\begin{center} 
		\includegraphics[width=17cm]{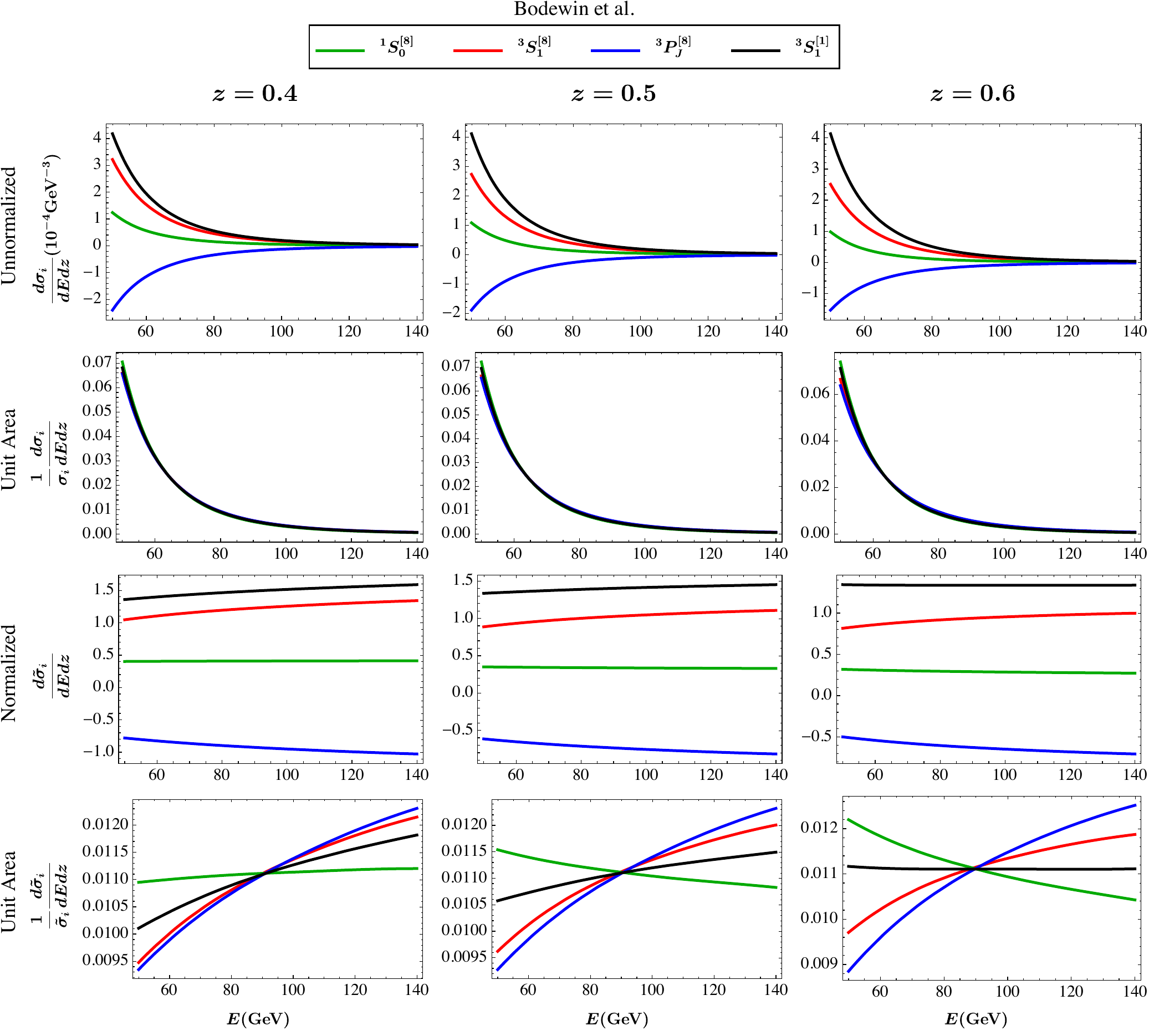}
		\end{center} 
		\vspace{-0.3cm}
		\caption{\baselineskip 3.0ex Unnormalized and normalized cross sections for Bodwin et al. extractions~\cite{Bodwin:2014gia}. The conventions followed are same as in Fig.~(\ref{plotCompNorm}).}
		\label{plotCompNormBO}
\end{figure*}
\clearpage
\section{Unnormalized and Normalized cross sections for Chao et.al}
\label{Chao} 
Fig.~(\ref{plotCompNormCH}) shows the unnormalized (Eq.~(\ref{eq:master})) and normalized cross section (Eq.~(\ref{eq:norm})) for Chao et al.'s LDME extractions~\cite{Chao:2012iv}. Similar to Bodwin et al., these LDMEs result in a cancellation between the $^3S_{1}^{[8]}$ and $^3P_{J}^{[8]}$ channels.
\begin{figure*}[htbp!]
	\begin{center} 
		\includegraphics[width=17cm]{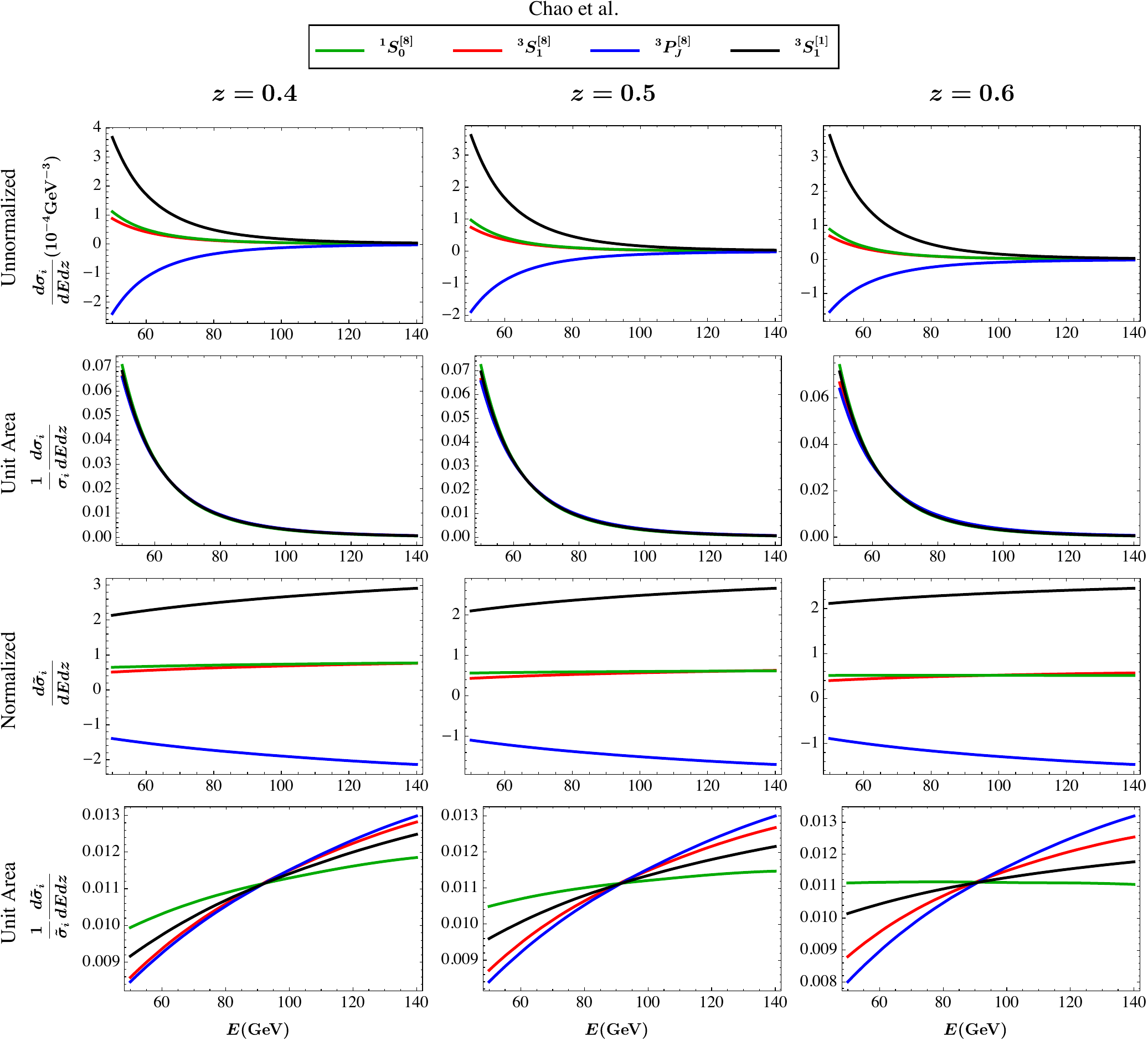}
		\end{center} 
		\vspace{-0.3cm}
		\caption{\baselineskip 3.0ex Unnormalized and normalized cross sections for Chao et al. extractions~\cite{Chao:2012iv}. The conventions followed are same as in Fig.~(\ref{plotCompNorm}).}
		\label{plotCompNormCH}
\end{figure*}
\clearpage
\section{Insensitivity to $z_{min}$ and $z_{max}$}
\label{NormalizationInsensitivity} 
Comparison of the normalized cross sections (Eq.~(\ref{eq:norm})) for different values of $z_{min}$ and $z_{max}$ is shown. This confirms that the discussion in section~\ref{sec:ldmedependent} is not sensitive to $(z_{min},z_{max})$ since the shapes of different LDMEs do not change. For validity of the factorization formula Eq.~(\ref{eq:master}), we don't pick $z_{min}$ too close to $0$ and $z_{max}$ too close to $1$.
\begin{figure*}[htbp!]
	\begin{center} 
\includegraphics[width=17cm]{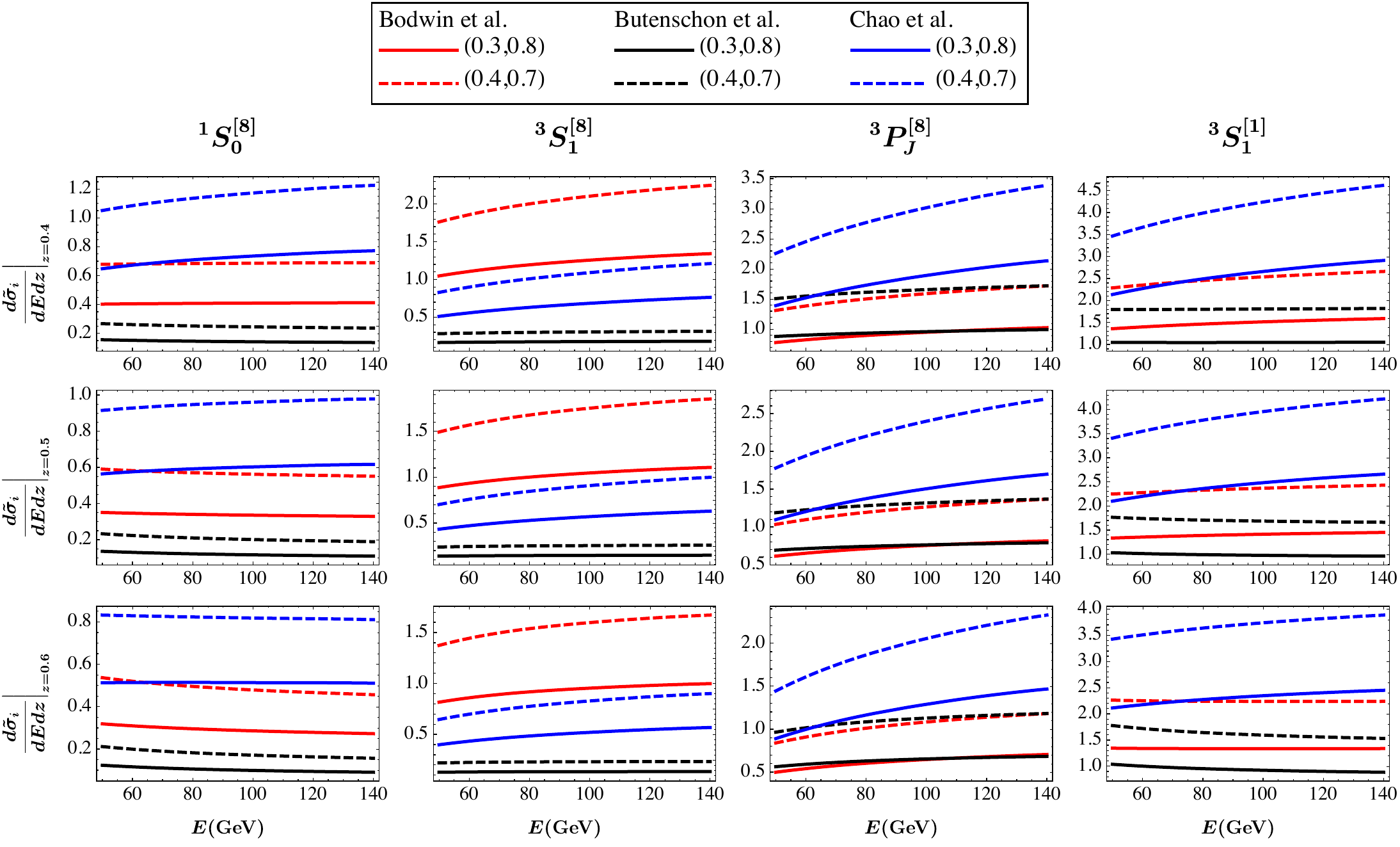}
\end{center}3
\vspace{-0.3cm}
\caption{\baselineskip 3.0ex Solid curves correspond to $(z_{min},z_{max})=(0.3,0.8)$ and the dashed curves $(z_{min},z_{max})=(0.4,0.7)$. Due to the change in normalization, all the curves shift upwards without changing their qualitative shapes.}
\label{plotNormVar}
\end{figure*}
\clearpage
\section{Normalization using only color octet channels}
\label{ColorOctet} 
Fig.~(\ref{plotOctNorm}) shows the cross section for the different $J/\psi$ production channels based on the LDMEs in Ref.~\cite{Bodwin:2014gia} and Ref.~\cite{Butenschoen:2011yh,Butenschoen:2012qr} with the contribution of $^3S_{1}^{[1]}$ channel ignored in Eq.~(\ref{eq:norm}), i.e., setting $\langle {\cal O}^{J/\psi}(^3S_1^{[1]}) \rangle$ to $0$. Since $^1S_0^{[8]}$ channel (green curves) has very different slopes for the two LDMEs, if the $^1S_0^{[8]}$ channel dominates at high $p_T$, then one can distinguish between these two extractions. We don't include Chao et al.'s extractions~\cite{Chao:2012iv} because it gives rise to a negative total cross section and so one can not ignore the color singlet contribution.
\begin{figure*}[htbp!]
	\begin{center} 
		\includegraphics[width=17cm]{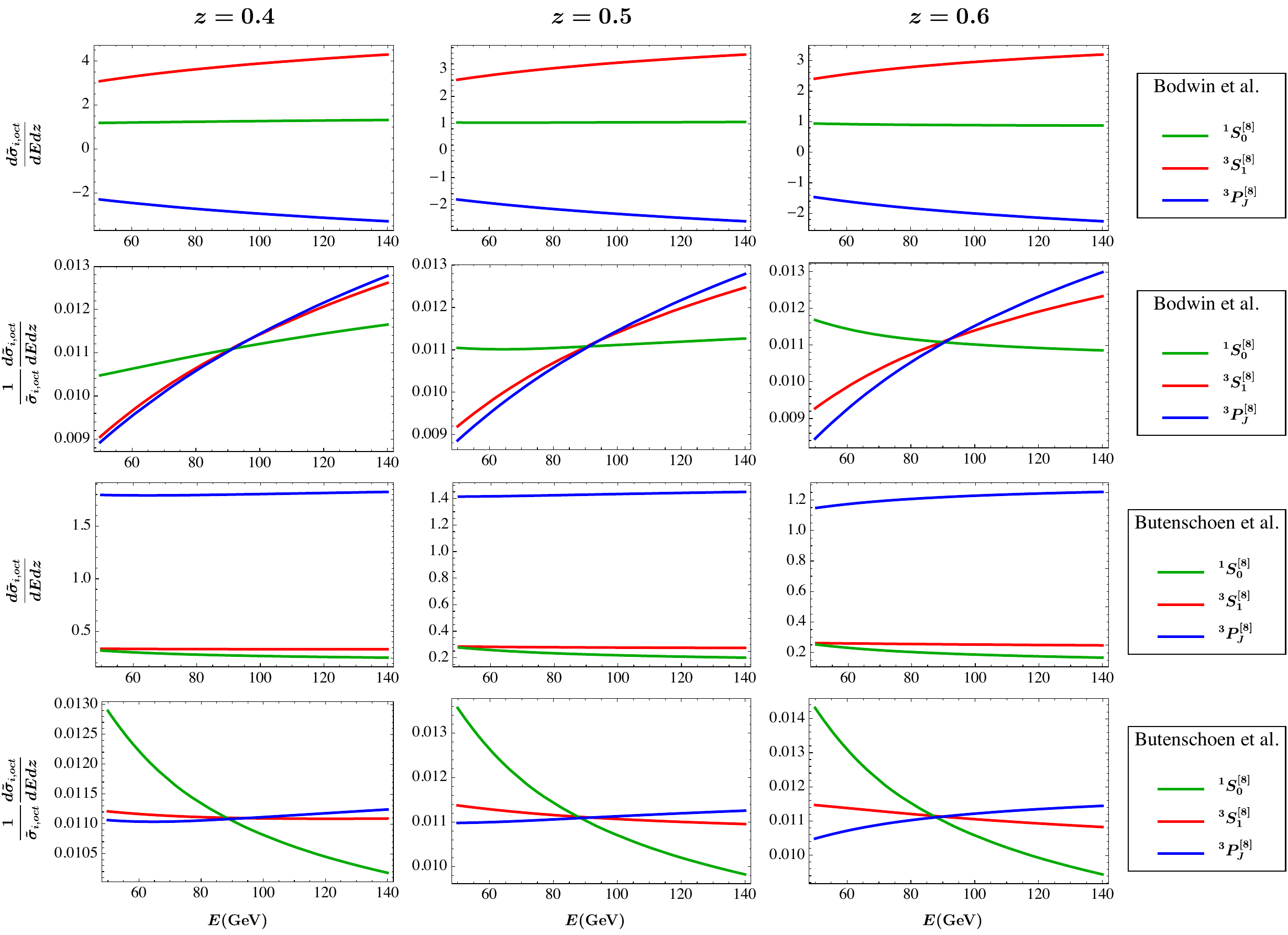}
	\end{center} 
	\vspace{-0.3cm}
	\caption{\baselineskip 3.0ex Cross section normalized by ignoring the $^3S_{1}^{[1]}$ channel contribution in Eq.~\ref{eq:norm}. The second and fourth row are obtained by normalizing the curves in the first and third row to unit area respectively.}
	\label{plotOctNorm}
\end{figure*}
\clearpage
\section{Lower $z$ plots}
\label{lowz} 
Fig.~(\ref{plotJetNorm234}) shows the $J/\psi$ production cross section (Eq.~(\ref{eq:normjc})) at lower $z$ values for all the three LDME extractions \cite{Bodwin:2014gia,Butenschoen:2011yh,Butenschoen:2012qr,Chao:2012iv} used in this paper.
\begin{figure*}[htbp!]
	\begin{center}
\includegraphics[width=17cm]{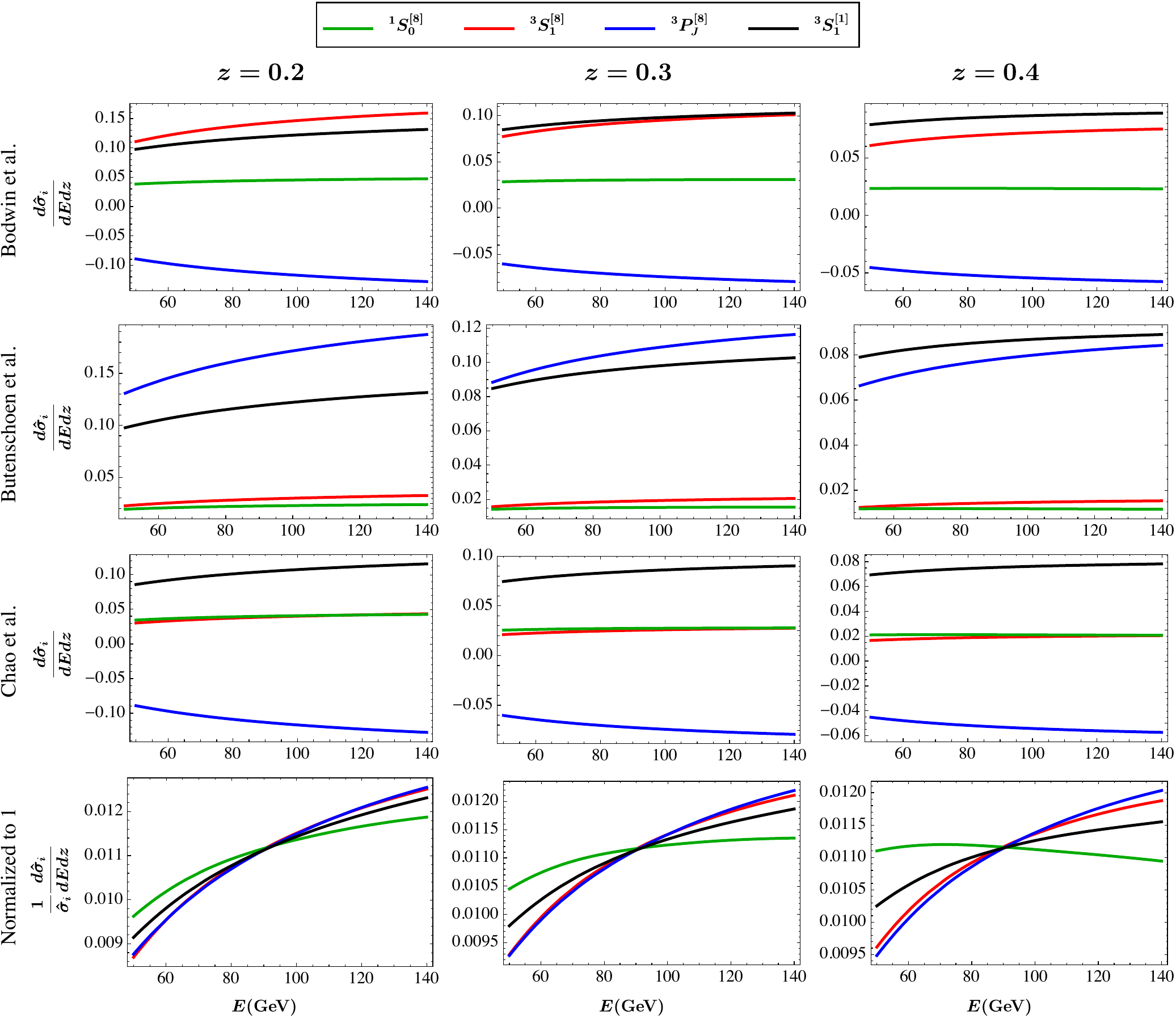}
	\end{center}
\vspace{-0.3cm}
\caption{\baselineskip 3.0ex Lower $z$ plots for the cross section (Eq.~(\ref{eq:normjc})). The conventions followed are same as those in Fig.~(\ref{plotJetNorm}).}
\label{plotJetNorm234}
\end{figure*}

}

\clearpage
\bibliographystyle{apsrev4-1}
\bibliography{paperRef}

\begin{thebibliography}{33}%
\makeatletter
\providecommand \@ifxundefined [1]{%
 \@ifx{#1\undefined}
}%
\providecommand \@ifnum [1]{%
 \ifnum #1\expandafter \@firstoftwo
 \else \expandafter \@secondoftwo
 \fi
}%
\providecommand \@ifx [1]{%
 \ifx #1\expandafter \@firstoftwo
 \else \expandafter \@secondoftwo
 \fi
}%
\providecommand \natexlab [1]{#1}%
\providecommand \enquote  [1]{``#1''}%
\providecommand \bibnamefont  [1]{#1}%
\providecommand \bibfnamefont [1]{#1}%
\providecommand \citenamefont [1]{#1}%
\providecommand \href@noop [0]{\@secondoftwo}%
\providecommand \href [0]{\begingroup \@sanitize@url \@href}%
\providecommand \@href[1]{\@@startlink{#1}\@@href}%
\providecommand \@@href[1]{\endgroup#1\@@endlink}%
\providecommand \@sanitize@url [0]{\catcode `\\12\catcode `\$12\catcode
  `\&12\catcode `\#12\catcode `\^12\catcode `\_12\catcode `\%12\relax}%
\providecommand \@@startlink[1]{}%
\providecommand \@@endlink[0]{}%
\providecommand \url  [0]{\begingroup\@sanitize@url \@url }%
\providecommand \@url [1]{\endgroup\@href {#1}{\urlprefix }}%
\providecommand \urlprefix  [0]{URL }%
\providecommand \Eprint [0]{\href }%
\providecommand \doibase [0]{http://dx.doi.org/}%
\providecommand \selectlanguage [0]{\@gobble}%
\providecommand \bibinfo  [0]{\@secondoftwo}%
\providecommand \bibfield  [0]{\@secondoftwo}%
\providecommand \translation [1]{[#1]}%
\providecommand \BibitemOpen [0]{}%
\providecommand \bibitemStop [0]{}%
\providecommand \bibitemNoStop [0]{.\EOS\space}%
\providecommand \EOS [0]{\spacefactor3000\relax}%
\providecommand \BibitemShut  [1]{\csname bibitem#1\endcsname}%
\let\auto@bib@innerbib\@empty
\bibitem [{\citenamefont {Aaij}\ \emph {et~al.}(2017)\citenamefont {Aaij} \emph
  {et~al.}}]{Aaij:2017fak}%
  \BibitemOpen
  \bibfield  {author} {\bibinfo {author} {\bibfnamefont {R.}~\bibnamefont
  {Aaij}} \emph {et~al.} (\bibinfo {collaboration} {LHCb}),\ }\href {\doibase
  10.1103/PhysRevLett.118.192001} {\bibfield  {journal} {\bibinfo  {journal}
  {Phys. Rev. Lett.}\ }\textbf {\bibinfo {volume} {118}},\ \bibinfo {pages}
  {192001} (\bibinfo {year} {2017})},\ \Eprint
  {http://arxiv.org/abs/1701.05116} {arXiv:1701.05116 [hep-ex]} \BibitemShut
  {NoStop}%
\bibitem [{\citenamefont {Kang}\ \emph {et~al.}(2017)\citenamefont {Kang},
  \citenamefont {Qiu}, \citenamefont {Ringer}, \citenamefont {Xing},\ and\
  \citenamefont {Zhang}}]{Ringer:2017}%
  \BibitemOpen
  \bibfield  {author} {\bibinfo {author} {\bibfnamefont {Z.~B.}\ \bibnamefont
  {Kang}}, \bibinfo {author} {\bibfnamefont {J.~W.}\ \bibnamefont {Qiu}},
  \bibinfo {author} {\bibfnamefont {F.}~\bibnamefont {Ringer}}, \bibinfo
  {author} {\bibfnamefont {H.}~\bibnamefont {Xing}}, \ and\ \bibinfo {author}
  {\bibfnamefont {H.}~\bibnamefont {Zhang}},\ }\href {\doibase
  10.1103/PhysRevLett.119.032001} {\bibfield  {journal} {\bibinfo  {journal}
  {Phys. Rev. Lett.}\ }\textbf {\bibinfo {volume} {119}},\ \bibinfo {pages}
  {032001} (\bibinfo {year} {2017})},\ \Eprint
  {http://arxiv.org/abs/1702.03287} {arXiv:1702.03287 [hep-ph]} \BibitemShut
  {NoStop}%
\bibitem [{\citenamefont {Bain}\ \emph {et~al.}(2017)\citenamefont {Bain},
  \citenamefont {Dai}, \citenamefont {Leibovich}, \citenamefont {Makris},\ and\
  \citenamefont {Mehen}}]{Bain:2017wvk}%
  \BibitemOpen
  \bibfield  {author} {\bibinfo {author} {\bibfnamefont {R.}~\bibnamefont
  {Bain}}, \bibinfo {author} {\bibfnamefont {L.}~\bibnamefont {Dai}}, \bibinfo
  {author} {\bibfnamefont {A.~K.}\ \bibnamefont {Leibovich}}, \bibinfo {author}
  {\bibfnamefont {Y.}~\bibnamefont {Makris}}, \ and\ \bibinfo {author}
  {\bibfnamefont {T.}~\bibnamefont {Mehen}},\ }\href {\doibase
  10.1103/PhysRevLett.119.032002} {\bibfield  {journal} {\bibinfo  {journal}
  {Phys. Rev. Lett.}\ }\textbf {\bibinfo {volume} {119}},\ \bibinfo {pages}
  {032002} (\bibinfo {year} {2017})},\ \Eprint
  {http://arxiv.org/abs/1702.05525} {arXiv:1702.05525 [hep-ph]} \BibitemShut
  {NoStop}%
\bibitem [{\citenamefont {Belyaev}\ \emph {et~al.}(2017)\citenamefont
  {Belyaev}, \citenamefont {Berezhnoy}, \citenamefont {Likhoded},\ and\
  \citenamefont {Luchinsky}}]{Belyaev:2017lbo}%
  \BibitemOpen
  \bibfield  {author} {\bibinfo {author} {\bibfnamefont {I.}~\bibnamefont
  {Belyaev}}, \bibinfo {author} {\bibfnamefont {A.~V.}\ \bibnamefont
  {Berezhnoy}}, \bibinfo {author} {\bibfnamefont {A.~K.}\ \bibnamefont
  {Likhoded}}, \ and\ \bibinfo {author} {\bibfnamefont {A.~V.}\ \bibnamefont
  {Luchinsky}},\ }\href@noop {} {\  (\bibinfo {year} {2017})},\ \Eprint
  {http://arxiv.org/abs/1703.09081} {arXiv:1703.09081 [hep-ph]} \BibitemShut
  {NoStop}%
\bibitem [{\citenamefont {Bodwin}\ \emph {et~al.}(1995)\citenamefont {Bodwin},
  \citenamefont {Braaten},\ and\ \citenamefont {Lepage}}]{Bodwin:1995}%
  \BibitemOpen
  \bibfield  {author} {\bibinfo {author} {\bibfnamefont {G.~T.}\ \bibnamefont
  {Bodwin}}, \bibinfo {author} {\bibfnamefont {E.}~\bibnamefont {Braaten}}, \
  and\ \bibinfo {author} {\bibfnamefont {G.~P.}\ \bibnamefont {Lepage}},\
  }\href {\doibase 10.1103/PhysRevD.51.1125} {\bibfield  {journal} {\bibinfo
  {journal} {Phys. Rev.}\ }\textbf {\bibinfo {volume} {D51}},\ \bibinfo {pages}
  {1125} (\bibinfo {year} {1995})},\ \bibinfo {note} {[Erratum-ibid. D 55, 5853
  (1997)]},\ \Eprint {http://arxiv.org/abs/hep-ph/9407339}
  {arXiv:hep-ph/9407339 [hep-ph]} \BibitemShut {NoStop}%
\bibitem [{\citenamefont {Caswell}\ and\ \citenamefont
  {Lepage}(1986)}]{Caswell:1986}%
  \BibitemOpen
  \bibfield  {author} {\bibinfo {author} {\bibfnamefont {W.~E.}\ \bibnamefont
  {Caswell}}\ and\ \bibinfo {author} {\bibfnamefont {G.~P.}\ \bibnamefont
  {Lepage}},\ }\href {\doibase 10.1016/0370-2693(86)91297-9} {\bibfield
  {journal} {\bibinfo  {journal} {Phys.Lett.}\ }\textbf {\bibinfo {volume}
  {B167}},\ \bibinfo {pages} {437} (\bibinfo {year} {1986})}\BibitemShut
  {NoStop}%
\bibitem [{\citenamefont {Brambilla}\ \emph {et~al.}(2000)\citenamefont
  {Brambilla}, \citenamefont {Pineda}, \citenamefont {Soto},\ and\
  \citenamefont {Vairo}}]{Brambilla:2000}%
  \BibitemOpen
  \bibfield  {author} {\bibinfo {author} {\bibfnamefont {N.}~\bibnamefont
  {Brambilla}}, \bibinfo {author} {\bibfnamefont {A.}~\bibnamefont {Pineda}},
  \bibinfo {author} {\bibfnamefont {J.}~\bibnamefont {Soto}}, \ and\ \bibinfo
  {author} {\bibfnamefont {A.}~\bibnamefont {Vairo}},\ }\href {\doibase
  10.1016/S0550-3213(99)00693-8} {\bibfield  {journal} {\bibinfo  {journal}
  {Nucl.Phys.}\ }\textbf {\bibinfo {volume} {B566}},\ \bibinfo {pages} {275}
  (\bibinfo {year} {2000})},\ \Eprint {http://arxiv.org/abs/hep-ph/9907240}
  {arXiv:hep-ph/9907240 [hep-ph]} \BibitemShut {NoStop}%
\bibitem [{\citenamefont {Luke}\ \emph {et~al.}(2000)\citenamefont {Luke},
  \citenamefont {Manohar},\ and\ \citenamefont {Rothstein}}]{Luke:2000}%
  \BibitemOpen
  \bibfield  {author} {\bibinfo {author} {\bibfnamefont {M.~E.}\ \bibnamefont
  {Luke}}, \bibinfo {author} {\bibfnamefont {A.~V.}\ \bibnamefont {Manohar}}, \
  and\ \bibinfo {author} {\bibfnamefont {I.~Z.}\ \bibnamefont {Rothstein}},\
  }\href {\doibase 10.1103/PhysRevD.61.074025} {\bibfield  {journal} {\bibinfo
  {journal} {Phys.Rev.}\ }\textbf {\bibinfo {volume} {D61}},\ \bibinfo {pages}
  {074025} (\bibinfo {year} {2000})},\ \Eprint
  {http://arxiv.org/abs/hep-ph/9910209} {arXiv:hep-ph/9910209 [hep-ph]}
  \BibitemShut {NoStop}%
\bibitem [{\citenamefont {Rothstein}\ \emph {et~al.}()\citenamefont
  {Rothstein}, \citenamefont {Shrivastava},\ and\ \citenamefont
  {Stewart}}]{Rothstein:2017}%
  \BibitemOpen
  \bibfield  {author} {\bibinfo {author} {\bibfnamefont {I.~Z.}\ \bibnamefont
  {Rothstein}}, \bibinfo {author} {\bibfnamefont {P.}~\bibnamefont
  {Shrivastava}}, \ and\ \bibinfo {author} {\bibfnamefont {I.~W.}\ \bibnamefont
  {Stewart}},\ }\href@noop {} {\ }\bibinfo {note} {(Unpublished)}\BibitemShut
  {NoStop}%
\bibitem [{\citenamefont {Bodwin}\ \emph {et~al.}(2014)\citenamefont {Bodwin},
  \citenamefont {Chung}, \citenamefont {Kim},\ and\ \citenamefont
  {Lee}}]{Bodwin:2014gia}%
  \BibitemOpen
  \bibfield  {author} {\bibinfo {author} {\bibfnamefont {G.~T.}\ \bibnamefont
  {Bodwin}}, \bibinfo {author} {\bibfnamefont {H.~S.}\ \bibnamefont {Chung}},
  \bibinfo {author} {\bibfnamefont {U.-R.}\ \bibnamefont {Kim}}, \ and\
  \bibinfo {author} {\bibfnamefont {J.}~\bibnamefont {Lee}},\ }\href {\doibase
  10.1103/PhysRevLett.113.022001} {\bibfield  {journal} {\bibinfo  {journal}
  {Phys. Rev. Lett.}\ }\textbf {\bibinfo {volume} {113}},\ \bibinfo {pages}
  {022001} (\bibinfo {year} {2014})},\ \Eprint {http://arxiv.org/abs/1403.3612}
  {arXiv:1403.3612 [hep-ph]} \BibitemShut {NoStop}%
\bibitem [{\citenamefont {Butenschoen}\ and\ \citenamefont
  {Kniehl}(2011)}]{Butenschoen:2011yh}%
  \BibitemOpen
  \bibfield  {author} {\bibinfo {author} {\bibfnamefont {M.}~\bibnamefont
  {Butenschoen}}\ and\ \bibinfo {author} {\bibfnamefont {B.~A.}\ \bibnamefont
  {Kniehl}},\ }\href {\doibase 10.1103/PhysRevD.84.051501} {\bibfield
  {journal} {\bibinfo  {journal} {Phys. Rev.}\ }\textbf {\bibinfo {volume}
  {D84}},\ \bibinfo {pages} {051501} (\bibinfo {year} {2011})},\ \Eprint
  {http://arxiv.org/abs/1105.0820} {arXiv:1105.0820 [hep-ph]} \BibitemShut
  {NoStop}%
\bibitem [{\citenamefont {Butenschoen}\ and\ \citenamefont
  {Kniehl}(2013)}]{Butenschoen:2012qr}%
  \BibitemOpen
  \bibfield  {author} {\bibinfo {author} {\bibfnamefont {M.}~\bibnamefont
  {Butenschoen}}\ and\ \bibinfo {author} {\bibfnamefont {B.~A.}\ \bibnamefont
  {Kniehl}},\ }\href {\doibase 10.1142/S0217732313500272} {\bibfield  {journal}
  {\bibinfo  {journal} {Mod. Phys. Lett.}\ }\textbf {\bibinfo {volume} {A28}},\
  \bibinfo {pages} {1350027} (\bibinfo {year} {2013})},\ \Eprint
  {http://arxiv.org/abs/1212.2037} {arXiv:1212.2037} \BibitemShut {NoStop}%
\bibitem [{\citenamefont {Chao}\ \emph {et~al.}(2012)\citenamefont {Chao},
  \citenamefont {Ma}, \citenamefont {Shao}, \citenamefont {Wang},\ and\
  \citenamefont {Zhang}}]{Chao:2012iv}%
  \BibitemOpen
  \bibfield  {author} {\bibinfo {author} {\bibfnamefont {K.~T.}\ \bibnamefont
  {Chao}}, \bibinfo {author} {\bibfnamefont {Y.~Q.}\ \bibnamefont {Ma}},
  \bibinfo {author} {\bibfnamefont {H.~S.}\ \bibnamefont {Shao}}, \bibinfo
  {author} {\bibfnamefont {K.}~\bibnamefont {Wang}}, \ and\ \bibinfo {author}
  {\bibfnamefont {Y.~J.}\ \bibnamefont {Zhang}},\ }\href {\doibase
  10.1103/PhysRevLett.108.242004} {\bibfield  {journal} {\bibinfo  {journal}
  {Phys. Rev. Lett.}\ }\textbf {\bibinfo {volume} {108}},\ \bibinfo {pages}
  {242004} (\bibinfo {year} {2012})},\ \Eprint {http://arxiv.org/abs/1201.2675}
  {arXiv:1201.2675 [hep-ph]} \BibitemShut {NoStop}%
\bibitem [{\citenamefont {Fleming}\ \emph {et~al.}(2001)\citenamefont
  {Fleming}, \citenamefont {Leibovich},\ and\ \citenamefont
  {Rothstein}}]{Fleming:2000}%
  \BibitemOpen
  \bibfield  {author} {\bibinfo {author} {\bibfnamefont {S.}~\bibnamefont
  {Fleming}}, \bibinfo {author} {\bibfnamefont {A.~K.}\ \bibnamefont
  {Leibovich}}, \ and\ \bibinfo {author} {\bibfnamefont {I.~Z.}\ \bibnamefont
  {Rothstein}},\ }\href {\doibase 10.1103/PhysRevD.64.036002} {\bibfield
  {journal} {\bibinfo  {journal} {Phys.Rev.}\ }\textbf {\bibinfo {volume}
  {D64}},\ \bibinfo {pages} {036002} (\bibinfo {year} {2001})},\ \Eprint
  {http://arxiv.org/abs/hep-ph/0012062} {arXiv:hep-ph/0012062 [hep-ph]}
  \BibitemShut {NoStop}%
\bibitem [{\citenamefont {Cho}\ and\ \citenamefont {Wise}(1995)}]{Cho:1995}%
  \BibitemOpen
  \bibfield  {author} {\bibinfo {author} {\bibfnamefont {P.}~\bibnamefont
  {Cho}}\ and\ \bibinfo {author} {\bibfnamefont {M.~B.}\ \bibnamefont {Wise}},\
  }\href {\doibase 10.1016/0370-2693(94)01658-Y} {\bibfield  {journal}
  {\bibinfo  {journal} {Phys.Lett.}\ }\textbf {\bibinfo {volume} {B346}},\
  \bibinfo {pages} {129} (\bibinfo {year} {1995})},\ \Eprint
  {http://arxiv.org/abs/hep-ph/9411303} {arXiv:hep-ph/9411303 [hep-ph]}
  \BibitemShut {NoStop}%
\bibitem [{\citenamefont {Beneke}\ and\ \citenamefont
  {Rothstein}(1996{\natexlab{a}})}]{Beneke:1996}%
  \BibitemOpen
  \bibfield  {author} {\bibinfo {author} {\bibfnamefont {M.}~\bibnamefont
  {Beneke}}\ and\ \bibinfo {author} {\bibfnamefont {I.~Z.}\ \bibnamefont
  {Rothstein}},\ }\href {\doibase 10.1016/0370-2693(96)00030-5} {\bibfield
  {journal} {\bibinfo  {journal} {Phys.Lett.}\ }\textbf {\bibinfo {volume}
  {B372}},\ \bibinfo {pages} {157} (\bibinfo {year} {1996}{\natexlab{a}})},\
  \bibinfo {note} {[Erratum-ibid. B389, 769 (1996)]},\ \Eprint
  {http://arxiv.org/abs/hep-ph/9509375} {arXiv:hep-ph/9509375 [hep-ph]}
  \BibitemShut {NoStop}%
\bibitem [{\citenamefont {Beneke}\ and\ \citenamefont
  {Rothstein}(1996{\natexlab{b}})}]{Beneke2:1996}%
  \BibitemOpen
  \bibfield  {author} {\bibinfo {author} {\bibfnamefont {M.}~\bibnamefont
  {Beneke}}\ and\ \bibinfo {author} {\bibfnamefont {I.~Z.}\ \bibnamefont
  {Rothstein}},\ }\href {\doibase 10.1103/PhysRevD.54.2005} {\bibfield
  {journal} {\bibinfo  {journal} {Phys.Rev.}\ }\textbf {\bibinfo {volume}
  {D54}},\ \bibinfo {pages} {2005} (\bibinfo {year} {1996}{\natexlab{b}})},\
  \bibinfo {note} {[Erratum-ibid. D54, 7082 (1996)]},\ \Eprint
  {http://arxiv.org/abs/hep-ph/9603400} {arXiv:hep-ph/9603400 [hep-ph]}
  \BibitemShut {NoStop}%
\bibitem [{\citenamefont {Abulencia}\ \emph {et~al.}(2007)\citenamefont
  {Abulencia} \emph {et~al.}}]{Abulencia:2007us}%
  \BibitemOpen
  \bibfield  {author} {\bibinfo {author} {\bibfnamefont {A.}~\bibnamefont
  {Abulencia}} \emph {et~al.} (\bibinfo {collaboration} {CDF}),\ }\href
  {\doibase 10.1103/PhysRevLett.99.132001} {\bibfield  {journal} {\bibinfo
  {journal} {Phys. Rev. Lett.}\ }\textbf {\bibinfo {volume} {99}},\ \bibinfo
  {pages} {132001} (\bibinfo {year} {2007})},\ \Eprint
  {http://arxiv.org/abs/0704.0638} {arXiv:0704.0638 [hep-ex]} \BibitemShut
  {NoStop}%
\bibitem [{\citenamefont {Chatrchyan}\ \emph {et~al.}(2013)\citenamefont
  {Chatrchyan} \emph {et~al.}}]{Chatrchyan:2013cla}%
  \BibitemOpen
  \bibfield  {author} {\bibinfo {author} {\bibfnamefont {S.}~\bibnamefont
  {Chatrchyan}} \emph {et~al.} (\bibinfo {collaboration} {CMS}),\ }\href
  {\doibase 10.1016/j.physletb.2013.10.055} {\bibfield  {journal} {\bibinfo
  {journal} {Phys. Lett.}\ }\textbf {\bibinfo {volume} {B727}},\ \bibinfo
  {pages} {381} (\bibinfo {year} {2013})},\ \Eprint
  {http://arxiv.org/abs/1307.6070} {arXiv:1307.6070 [hep-ex]} \BibitemShut
  {NoStop}%
\bibitem [{\citenamefont {Aaij}\ \emph {et~al.}(2013)\citenamefont {Aaij} \emph
  {et~al.}}]{Aaij:2013nlm}%
  \BibitemOpen
  \bibfield  {author} {\bibinfo {author} {\bibfnamefont {R.}~\bibnamefont
  {Aaij}} \emph {et~al.} (\bibinfo {collaboration} {LHCb}),\ }\href {\doibase
  10.1140/epjc/s10052-013-2631-3} {\bibfield  {journal} {\bibinfo  {journal}
  {Eur. Phys. J.}\ }\textbf {\bibinfo {volume} {C73}},\ \bibinfo {pages} {2631}
  (\bibinfo {year} {2013})},\ \Eprint {http://arxiv.org/abs/1307.6379}
  {arXiv:1307.6379 [hep-ex]} \BibitemShut {NoStop}%
\bibitem [{\citenamefont {Brambilla~et al.}(2011)}]{Brambilla:2011}%
  \BibitemOpen
  \bibfield  {author} {\bibinfo {author} {\bibfnamefont {N.}~\bibnamefont
  {Brambilla~et al.}},\ }\href {\doibase 10.1140/epjc/s10052-010-1534-9}
  {\bibfield  {journal} {\bibinfo  {journal} {Eur.Phys.J.}\ }\textbf {\bibinfo
  {volume} {C71}},\ \bibinfo {pages} {1534} (\bibinfo {year} {2011})},\ \Eprint
  {http://arxiv.org/abs/1010.5827} {arXiv:1010.5827 [hep-ph]} \BibitemShut
  {NoStop}%
\bibitem [{\citenamefont {Baumgart}\ \emph {et~al.}(2014)\citenamefont
  {Baumgart}, \citenamefont {Leibovich}, \citenamefont {Mehen},\ and\
  \citenamefont {Rothstein}}]{Baumgart:2014}%
  \BibitemOpen
  \bibfield  {author} {\bibinfo {author} {\bibfnamefont {M.}~\bibnamefont
  {Baumgart}}, \bibinfo {author} {\bibfnamefont {A.~K.}\ \bibnamefont
  {Leibovich}}, \bibinfo {author} {\bibfnamefont {T.}~\bibnamefont {Mehen}}, \
  and\ \bibinfo {author} {\bibfnamefont {I.~Z.}\ \bibnamefont {Rothstein}},\
  }\href {\doibase 10.1007/JHEP11(2014)003} {\bibfield  {journal} {\bibinfo
  {journal} {JHEP}\ }\textbf {\bibinfo {volume} {1411}},\ \bibinfo {pages}
  {003} (\bibinfo {year} {2014})},\ \Eprint {http://arxiv.org/abs/1406.2295}
  {arXiv:1406.2295 [hep-ph]} \BibitemShut {NoStop}%
\bibitem [{\citenamefont {Procura}\ and\ \citenamefont
  {Stewart}(2010)}]{Stewart:2010}%
  \BibitemOpen
  \bibfield  {author} {\bibinfo {author} {\bibfnamefont {M.}~\bibnamefont
  {Procura}}\ and\ \bibinfo {author} {\bibfnamefont {I.~W.}\ \bibnamefont
  {Stewart}},\ }\href {\doibase 10.1103/PhysRevD.81.074009} {\bibfield
  {journal} {\bibinfo  {journal} {Phys. Rev.}\ }\textbf {\bibinfo {volume}
  {D81}},\ \bibinfo {pages} {074009} (\bibinfo {year} {2010})},\ \bibinfo
  {note} {[Erratum-ibid. D83, 039902 (2011)]},\ \Eprint
  {http://arxiv.org/abs/0911.4980} {arXiv:0911.4980 [hep-ph]} \BibitemShut
  {NoStop}%
\bibitem [{\citenamefont {Liu}(2011)}]{Liu:2011}%
  \BibitemOpen
  \bibfield  {author} {\bibinfo {author} {\bibfnamefont {X.}~\bibnamefont
  {Liu}},\ }\href {\doibase 10.1016/j.physletb.2011.03.055} {\bibfield
  {journal} {\bibinfo  {journal} {Phys.Lett.}\ }\textbf {\bibinfo {volume}
  {B699}},\ \bibinfo {pages} {87} (\bibinfo {year} {2011})},\ \Eprint
  {http://arxiv.org/abs/1011.3872} {arXiv:1011.3872 [hep-ph]} \BibitemShut
  {NoStop}%
\bibitem [{\citenamefont {Jain}\ \emph {et~al.}(2011)\citenamefont {Jain},
  \citenamefont {Procura},\ and\ \citenamefont {Waalewijn}}]{Jain:2011}%
  \BibitemOpen
  \bibfield  {author} {\bibinfo {author} {\bibfnamefont {A.}~\bibnamefont
  {Jain}}, \bibinfo {author} {\bibfnamefont {M.}~\bibnamefont {Procura}}, \
  and\ \bibinfo {author} {\bibfnamefont {W.~J.}\ \bibnamefont {Waalewijn}},\
  }\href {\doibase 10.1007/JHEP05(2011)035} {\bibfield  {journal} {\bibinfo
  {journal} {JHEP}\ }\textbf {\bibinfo {volume} {1105}},\ \bibinfo {pages}
  {035} (\bibinfo {year} {2011})},\ \Eprint {http://arxiv.org/abs/1101.4953}
  {arXiv:1101.4953 [hep-ph]} \BibitemShut {NoStop}%
\bibitem [{\citenamefont {Jain}\ \emph {et~al.}(2012)\citenamefont {Jain},
  \citenamefont {Procura},\ and\ \citenamefont {Waalewijn}}]{Jain:2012}%
  \BibitemOpen
  \bibfield  {author} {\bibinfo {author} {\bibfnamefont {A.}~\bibnamefont
  {Jain}}, \bibinfo {author} {\bibfnamefont {M.}~\bibnamefont {Procura}}, \
  and\ \bibinfo {author} {\bibfnamefont {W.~J.}\ \bibnamefont {Waalewijn}},\
  }\href {\doibase 10.1007/JHEP04(2012)132} {\bibfield  {journal} {\bibinfo
  {journal} {JHEP}\ }\textbf {\bibinfo {volume} {1204}},\ \bibinfo {pages}
  {132} (\bibinfo {year} {2012})},\ \Eprint {http://arxiv.org/abs/1110.0839}
  {arXiv:1110.0839 [hep-ph]} \BibitemShut {NoStop}%
\bibitem [{\citenamefont {Procura}\ and\ \citenamefont
  {Waalewijn}(2012)}]{Procura:2012}%
  \BibitemOpen
  \bibfield  {author} {\bibinfo {author} {\bibfnamefont {M.}~\bibnamefont
  {Procura}}\ and\ \bibinfo {author} {\bibfnamefont {W.~J.}\ \bibnamefont
  {Waalewijn}},\ }\href {\doibase 10.1103/PhysRevD.85.114041} {\bibfield
  {journal} {\bibinfo  {journal} {Phys. Rev.}\ }\textbf {\bibinfo {volume}
  {D85}},\ \bibinfo {pages} {114041} (\bibinfo {year} {2012})},\ \Eprint
  {http://arxiv.org/abs/1111.6605} {arXiv:1111.6605 [hep-ph]} \BibitemShut
  {NoStop}%
\bibitem [{\citenamefont {Jain}\ \emph {et~al.}(2013)\citenamefont {Jain},
  \citenamefont {Procura}, \citenamefont {Shotwell},\ and\ \citenamefont
  {Waalewijn}}]{Jain:2013}%
  \BibitemOpen
  \bibfield  {author} {\bibinfo {author} {\bibfnamefont {A.}~\bibnamefont
  {Jain}}, \bibinfo {author} {\bibfnamefont {M.}~\bibnamefont {Procura}},
  \bibinfo {author} {\bibfnamefont {B.}~\bibnamefont {Shotwell}}, \ and\
  \bibinfo {author} {\bibfnamefont {W.~J.}\ \bibnamefont {Waalewijn}},\ }\href
  {\doibase 10.1103/PhysRevD.87.074013} {\bibfield  {journal} {\bibinfo
  {journal} {Phys.Rev.}\ }\textbf {\bibinfo {volume} {D87}},\ \bibinfo {pages}
  {074013} (\bibinfo {year} {2013})},\ \bibinfo {note} {no.7},\ \Eprint
  {http://arxiv.org/abs/1207.4788} {arXiv:1207.4788 [hep-ph]} \BibitemShut
  {NoStop}%
\bibitem [{\citenamefont {Bauer}\ and\ \citenamefont
  {Mereghetti}()}]{Bauer:2013}%
  \BibitemOpen
  \bibfield  {author} {\bibinfo {author} {\bibfnamefont {C.~W.}\ \bibnamefont
  {Bauer}}\ and\ \bibinfo {author} {\bibfnamefont {E.}~\bibnamefont
  {Mereghetti}},\ }\href@noop {} {\ }\Eprint {http://arxiv.org/abs/1312.5605}
  {arXiv:1312.5605 [hep-ph]} \BibitemShut {NoStop}%
\bibitem [{\citenamefont {Ellis}\ \emph {et~al.}(2010)\citenamefont {Ellis},
  \citenamefont {Hornig}, \citenamefont {Lee}, \citenamefont {Vermilion},\ and\
  \citenamefont {Walsh}}]{ChrisLee:2010}%
  \BibitemOpen
  \bibfield  {author} {\bibinfo {author} {\bibfnamefont {S.~D.}\ \bibnamefont
  {Ellis}}, \bibinfo {author} {\bibfnamefont {A.}~\bibnamefont {Hornig}},
  \bibinfo {author} {\bibfnamefont {C.}~\bibnamefont {Lee}}, \bibinfo {author}
  {\bibfnamefont {C.~K.}\ \bibnamefont {Vermilion}}, \ and\ \bibinfo {author}
  {\bibfnamefont {J.~R.}\ \bibnamefont {Walsh}},\ }\href {\doibase
  10.1007/JHEP11(2010)101} {\bibfield  {journal} {\bibinfo  {journal} {JHEP}\
  }\textbf {\bibinfo {volume} {1011}},\ \bibinfo {pages} {101} (\bibinfo {year}
  {2010})},\ \Eprint {http://arxiv.org/abs/1001.0014} {arXiv:1001.0014
  [hep-ph]} \BibitemShut {NoStop}%
\bibitem [{\citenamefont {Ellis}\ \emph {et~al.}(2003)\citenamefont {Ellis},
  \citenamefont {Stirling},\ and\ \citenamefont {Webber}}]{Collider}%
  \BibitemOpen
  \bibfield  {author} {\bibinfo {author} {\bibfnamefont {R.~K.}\ \bibnamefont
  {Ellis}}, \bibinfo {author} {\bibfnamefont {W.~J.}\ \bibnamefont {Stirling}},
  \ and\ \bibinfo {author} {\bibfnamefont {B.~R.}\ \bibnamefont {Webber}},\
  }\href@noop {} {\emph {\bibinfo {title} {QCD and Collider Physics}}}\
  (\bibinfo  {publisher} {Cambridge University Press},\ \bibinfo {year}
  {2003})\BibitemShut {NoStop}%
\bibitem [{\citenamefont {Martin}\ \emph {et~al.}(2009)\citenamefont {Martin},
  \citenamefont {Stirling}, \citenamefont {Throne},\ and\ \citenamefont
  {Watt}}]{MSTW:2008nlo}%
  \BibitemOpen
  \bibfield  {author} {\bibinfo {author} {\bibfnamefont {A.~D.}\ \bibnamefont
  {Martin}}, \bibinfo {author} {\bibfnamefont {W.~J.}\ \bibnamefont
  {Stirling}}, \bibinfo {author} {\bibfnamefont {R.~S.}\ \bibnamefont
  {Throne}}, \ and\ \bibinfo {author} {\bibfnamefont {G.}~\bibnamefont
  {Watt}},\ }\href {\doibase 10.1140/epjc/s10052-009-1072-5} {\bibfield
  {journal} {\bibinfo  {journal} {Eur.Phys.J.}\ }\textbf {\bibinfo {volume}
  {C63}},\ \bibinfo {pages} {189} (\bibinfo {year} {2009})},\ \Eprint
  {http://arxiv.org/abs/0901.0002} {arXiv:0901.0002 [hep-ph]} \BibitemShut
  {NoStop}%
\bibitem [{\citenamefont {Bodwin}()}]{Bodwin:private}%
  \BibitemOpen
  \bibfield  {author} {\bibinfo {author} {\bibfnamefont {G.~T.}\ \bibnamefont
  {Bodwin}},\ }\href@noop {} {\bibinfo  {journal} {private communication}\
  }\BibitemShut {NoStop}%
\end{thebibliography}%


\end{document}